\documentstyle[preprint,tighten,aps,epsfig]{revtex}
\begin{document}

\title{$\Lambda NN$ and $\Sigma NN$ systems at threshold}

\author{H. Garcilazo$^{(1,3)}$, T. Fern\'andez-Caram\'es$^{(2)}$, 
and A. Valcarce$^{(3)}$}
\address{$(1)$ Escuela Superior de F\'\i sica y Matem\'aticas, 
Instituto Polit\'ecnico Nacional, \\ 
Edificio 9, 07738 M\'exico D.F., Mexico}
\address{$(2)$ Departamento de F\' \i sica Te\'orica e IFIC, 
Universidad de Valencia - CSIC, \\ 
E-46100 Burjassot, Valencia, Spain}
\address{$(3)$ Departamento de F\'\i sica Fundamental, 
Universidad de Salamanca,\\
E-37008 Salamanca, Spain}
\maketitle

\date{\today}
\begin{abstract}
We calculate the hypertriton binding energy and the
$\Lambda d$ and $\Sigma d$ scattering lengths using
baryon-baryon interactions obtained from a chiral constituent
quark model. We study consistently the $\Lambda NN$ 
and $\Sigma NN$ systems analyzing the effect of the
$\Sigma \leftrightarrow \Lambda$ conversion. 
Our interactions correctly predict the hypertriton binding
energy. The $(I,J)=(0,3/2)$ $\Lambda NN$ channel is also attractive and
it might have a bound state. From the condition of
nonexistence of a (0,3/2) $\Lambda NN$ bound state, an
upper limit for the spin-triplet $\Lambda N$ scattering length
is obtained. We also present results for the elastic and
inelastic $\Sigma N$ and $\Lambda N$ cross sections.
The consistent description of the $\Sigma N$ scattering cross
sections imposes a lower limit for the corresponding
spin-triplet scattering lengths.
In the $\Sigma NN$ system the only attractive 
channels are $(I,J)=(1,1/2)$ and $(0,1/2)$, the
$(1,1/2)$ state being the most attractive one.
\end{abstract}

\maketitle

\section{Introduction}

The deuteron plays an important role in conventional nonstrange
nuclear physics by constraining our models of the nucleon-nucleon ($NN$)
force. This is a consequence of the precision with which we can 
measure properties of a bound
system, exceeding by far that possible in measurements of the 
scattering amplitudes. Because neither the $\Lambda N$ nor the
$\Sigma N$ (spin-triplet or spin-singlet) interactions possess
sufficient strength to support a bound state, it is the
hypertriton that plays the important role
of the deuteron in hypernuclear physics. It is therefore the ground
state of the $\Lambda NN$ system $(J^P=1/2^+,I=0)$, that must be 
used to constrain our models of the hyperon-nucleon ($YN$) force.
In fact, the rather small binding energy of the hypertriton turned out to be elusive
for different meson-theoretical interactions \cite{Miy93,Nog02}. 

Unfortunately, there is hardly any scattering data available for the 
hyperon-nucleon ($YN$) system. The sparse database for $\Lambda N$ and 
$\Sigma N$ scattering and reactions is inadequate to fully determine 
the $YN$ interactions. Therefore, interaction models are generally
constrained by flavor $SU(3)$ symmetry. In this spirit, one meson-exchange
\cite{Rij99} and quark-model based \cite{Fuj06} forces have been developed 
which are all consistent with the scarce $YN$ database.

The chiral constituent quark model has been very successful in the
simultaneous description of the baryon-baryon interaction and the baryon 
spectrum as well as in the study of the two- and
three-baryon bound-state problem for the nonstrange sector \cite{Val05}.
A simple generalization of this model to the strange sector has also been
recently applied to study the meson and baryon spectra \cite{Gar05} and the 
$\Sigma NN$ bound-state problem \cite{Ter06}.
In this last work we solved the Faddeev equations of the coupled 
$\Sigma NN - \Lambda NN$ system below the $\Sigma d$ threshold in
order to study the amount of attraction present in the different
isospin-spin channels $(I,J)$. 

In this work, we will continue our study of the strangeness $\hat S=-1$
three-baryon systems by considering also the region of $\Lambda NN$ bound
states, where the hypertriton lies, and calculating for the first
time the $\Lambda d$ and $\Sigma d$ scattering lengths.
In the past the question was posed whether other states 
than that of the hypertriton, and in particular the $(I,J)=(0,3/2)$,
could also be bound. We shall investigate this question by studying 
simultaneously all $\Lambda NN$ and $\Sigma NN$ states with $J=1/2,3/2$
and $I=0,1,2$. We will also ask not only for the description of the elastic 
$\Lambda N$ cross section, but also for the elastic and inelastic 
$\Sigma N$ cross sections.
Therefore we pursue a simultaneous description of almost all
experimentally available observables in the two- and three-baryon
systems with strangeness $\hat S=-1$. The set of observables used would put
stringent conditions to any potential model approach. In particular,
this study will provide with a powerful tool to determine the relative 
strength of the spin-singlet and spin-triplet
hyperon-nucleon interactions, allowing to refine details on the
model interaction. 

The paper is organized as follows. 
Our formalism for the baryon-baryon interactions and the Faddeev
equations for the bound-state problem of the coupled $\Sigma NN - \Lambda NN$
system has been already presented in Ref. \cite{Ter06}. Thus in the next
section we will resume some relevant aspects of the interacting potential
as well as we will present the corresponding equations 
for $\Lambda d$ and $\Sigma d$
scattering at threshold and the expressions for the scattering lengths.
Section \ref{res} will be devoted to present and to discuss the results
obtained. Finally, in section \ref{sum} we will resume our most relevant
conclusions.

\section{Formalism}

\subsection{The two-body interactions}

The baryon-baryon interactions involved in the study of the coupled
$\Sigma NN - \Lambda NN$ system are obtained from the chiral 
constituent quark model \cite{Val05,Gar05}. In this model baryons
are described as clusters of three interacting massive (constituent) quarks,
the mass coming from the spontaneous breaking of chiral symmetry. The
first ingredient of the quark-quark interaction is a confining
potential ($CON$). Perturbative aspects of QCD are taken into account
by means of a one-gluon potential ($OGE$). Spontaneous breaking of 
chiral symmetry gives rise to boson exchanges
between quarks. In particular, there appear pseudoscalar 
boson exchanges, that have already been considered in our model \cite{Ter06},
and their corresponding scalar partners: $\sigma$, $\kappa$, $f_0$ 
and $a_0$. In our previous study of the baryon-baryon interaction for
systems with strangeness, the scalar meson-exchange
potential was used as in the nonstrange
systems, i.e. only $\sigma$-meson exchange. 
Here we first analyze the effect of the inclusion of the
other scalars of the octet.  These scalar potentials have 
the same functional form and a different $SU(3)$ operatorial dependence, i.e.,

\begin{equation}
V ({\vec r}_{ij}) = 
V_{a_0} ({\vec r}_{ij}) \sum_{F=1}^3 \lambda_i^F \cdot \lambda_j^F +
V_\kappa ({\vec r}_{ij}) \sum_{F=4}^7 \lambda_i^F \cdot \lambda_j^F +
V_{f_0} ({\vec r}_{ij}) \lambda_i^8 \cdot \lambda_j^8 +
V_{\sigma} ({\vec r}_{ij}) \lambda_i^0 \cdot \lambda_j^0 
\label{eq1}
\end{equation}

\noindent
where

\begin{equation}
V_{k} ({\vec r}_{ij}) = - {g_{ch} \over {4 \, \pi}} \,
{\Lambda_{k}^2 \over \Lambda_{k}^2 - m_{k}^2}
\, m_{k} \, \left[ Y (m_{k} \, r_{ij})-
{\Lambda_{k} \over {m_{k}}} \,
Y (\Lambda_{k} \, r_{ij}) \right] \, ,
\label{OSE}
\end{equation}

\noindent
with $k=a_0, \kappa, f_0 $ or $\sigma$, and $g_{ch}$ 
is the quark-meson coupling constant \cite{Val05}.
We give in Table \ref{t0}
the expectation value of the different flavor operators appearing
in Eq. (\ref{eq1}) for the different two-body systems
involved in our problem. If one takes additionally into account the
smallness of the scalar mixing angle \cite{Yao06}, one concludes
that the $f_0$ meson does not contribute.
Thus, there are two novelties when 
considering the full $SU(3)$ scalar octet.
Firstly, a reduction in the strength of the $\sigma -$meson exchange
potential, and second an increase of the attraction
for spin-singlet and a decrease for spin-triplet channels mainly due to 
the $\kappa -$meson exchange.
As a consequence, the effect of the scalar octet can be effectively
taken into account by using a single scalar exchange
of the form \cite{Ter07},

\begin{equation}
V_{s} ({\vec r}_{ij}) = - {g_{ch} \over {4 \, \pi}} \,
{\Lambda_{s}^2 \over \Lambda_{s}^2 - m_{s}^2}
\, m_{s} \, \left[ Y (m_{s} \, r_{ij})-
{\Lambda_{s} \over {m_{s}}} \,
Y (\Lambda_{s} \, r_{ij}) \right] \, ,
\label{sca}
\end{equation}

\noindent
with different parametrization for spin-singlet and spin-triplet
channels. A parametrization of the full scalar octet exchange
potential is obtained using the parameters given in Table \ref{t0b}.

Finally, the quark-quark interaction will read: 
\begin{equation}
V_{qq}(\vec{r}_{ij})=V_{CON}(\vec{r}_{ij})+V_{OGE}(\vec{r}_{ij})+V_{\chi}
(\vec{r}_{ij})+V_{s}(\vec{r}_{ij}) \,\,,
\label{int}
\end{equation}%
where the $i$ and $j$ indices are associated with $i$ and $j$ quarks
respectively, ${\vec{r}}_{ij}$ stands for the interquark distance, 
$V_{\chi}$ denotes the pseudoscalar meson-exchange interactions
discussed in Refs. \cite{Gar05,Ter06}, and $V_s$ stands for the 
effective scalar meson-exchange potential just described.
Explicit expression of all the
interacting potentials and a more detailed discussion 
of the model can be found in Ref. \cite{Gar05}.
In order to derive the local $NB_1\to NB_2$ interactions from the
basic $qq$ interaction defined above we use a Born-Oppenheimer
approximation. Explicitly, the potential is calculated as follows,

\begin{equation}
V_{NB_1 (L \, S \, T) \rightarrow NB_2 (L^{\prime}\, S^{\prime}\, T)} (R) =
\xi_{L \,S \, T}^{L^{\prime}\, S^{\prime}\, T} (R) \, - \, \xi_{L \,S \,
T}^{L^{\prime}\, S^{\prime}\, T} (\infty) \, ,  \label{Poten1}
\end{equation}

\noindent where

\begin{equation}
\xi_{L \, S \, T}^{L^{\prime}\, S^{\prime}\, T} (R) \, = \, {\frac{{\left
\langle \Psi_{NB_2 }^{L^{\prime}\, S^{\prime}\, T} ({\vec R}) \mid
\sum_{i<j=1}^{6} V_{qq}({\vec r}_{ij}) \mid \Psi_{NB_1 }^{L \, S \, T} ({\vec R%
}) \right \rangle} }{{\sqrt{\left \langle \Psi_{NB_2 }^{L^{\prime}\,
S^{\prime}\, T} ({\vec R}) \mid \Psi_{NB_2 }^{L^{\prime}\, S^{\prime}\, T} ({%
\vec R}) \right \rangle} \sqrt{\left \langle \Psi_{NB_1 }^{L \, S \, T} ({\vec %
R}) \mid \Psi_{NB_1 }^{L \, S \, T} ({\vec R}) \right \rangle}}}} \, .
\label{Poten2}
\end{equation}
In the last expression the quark coordinates are integrated out keeping $R$
fixed, the resulting interaction being a function of the $N-B_i$ 
relative distance. The
wave function $\Psi_{NB_i}^{L \, S \, T}({\vec R})$ for the two-baryon system
is discussed in detail in Ref. \cite{Val05}.

\subsection{$\Sigma d$ and $\Lambda d$ scattering at threshold}

Our method \cite{Ter06} to transform the Faddeev equations from being integral 
equations in two continuous variables into integral equations in just one
continuous variable is based in the expansion of the two-body $t-$matrices 
\begin{equation}
t_{i;s_ii_i}(p_i,p^\prime_i;e)=\sum_{nr}P_n(x_i)
\tau_{i;s_ii_i}^{nr}(e)P_r(x^\prime_i),
\label{for11}
\end{equation}
where $P_n$ and $P_r$ are Legendre polynomials,

\begin{equation}
x_{i}={\frac{p_{i}-b}{p_{i}+b}},  \label{for9}
\end{equation}%

\begin{equation}
x_{i}^\prime={\frac{p_{i}^\prime-b}{p_{i}^\prime+b}},  \label{for9p}
\end{equation}%
and $p_i$ and $p_i^\prime$ are the initial and final relative momenta of
the pair $jk$ while $b$ is a scale parameter of which the results do not
depend on.

Following the same notation as in Ref. \cite{Ter06}, where particle 1 is
the hyperon and particles 2 and 3 are the two nucleons, the integral
equations for $\beta d$ scattering at threshold with $\beta=\Sigma$
or $\Lambda$ are in the case of pure $S-$wave configurations,

\begin{eqnarray}
T_{2;SI;\beta}^{ns_2i_2}(q_2) & = & 
B_{2;SI;\beta}^{ns_2i_2}(q_2)  
+\sum_{ms_3i_3} \int_0^\infty dq_3\,\left[
(-1)^{1+\sigma _{1}+\sigma _{3}-s_{2}+\tau _{1}+\tau _{3}-i_{2}} 
A_{23;SI}^{ns_2i_2ms_3i_3}(q_2,q_3;E) \right.
\nonumber \\
& + &  \left. 2\sum_{rs_1i_1} \int_0^\infty dq_1\,
A_{31;SI}^{ns_2i_2rs_1i_1}(q_2,q_1;E)
A_{13;SI}^{rs_1i_1ms_3i_3}(q_1,q_3;E) \right]
T_{2;SI}^{ms_3i_3}(q_3),
\label{for20}
\end{eqnarray}
where $\sigma _{1}$ ($\tau _{1})$and $\sigma _{3}$ ($\tau _{3})$ stand for
the spin (isospin) of the hyperon and the nucleon respectively while
$s_i$ and $i_i$ are the spin and isospin of the pair $jk$.
$T_{2;SI;\beta}^{ns_2i_2}(q_2)$ is a two-component vector

\begin{equation}
T_{2;SI;\beta}^{ns_2i_2}(q_2) = \left( \matrix{
T_{2;SI;\Sigma\beta}^{ns_2i_2}(q_2)  \cr
T_{2;SI;\Lambda\beta}^{ns_2i_2}(q_2) \cr } \right),
\label{for2}
\end{equation}
while the kernel of Eq. (\ref{for20}) is a $2\times 2$ matrix defined by

\begin{equation}
A_{23;SI}^{ns_2i_2ms_3i_3}(q_2,q_3;E)=\left(\matrix{
A_{23;SI;\Sigma\Sigma}^{ns_2i_2ms_3i_3}(q_2,q_3;E)&
A_{23;SI;\Sigma\Lambda}^{ns_2i_2ms_3i_3}(q_2,q_3;E)\cr
A_{23;SI;\Lambda\Sigma}^{ns_2i_2ms_3i_3}(q_2,q_3;E)&
A_{23;SI;\Lambda\Lambda}^{ns_2i_2ms_3i_3}(q_2,q_3;E)\cr }\right),
\label{gl2}
\end{equation}

\begin{equation}
A_{31;SI}^{ns_2i_2rs_1i_1}(q_2,q_1;E)=\left(\matrix{
A_{31;SI;\Sigma N(\Sigma)}^{ns_2i_2rs_1i_1}(q_2,q_1;E)&
A_{31;SI;\Sigma N(\Lambda)}^{ns_2i_2rs_1i_1}(q_2,q_1;E)\cr
A_{31;SI;\Lambda N(\Sigma)}^{ns_2i_2rs_1i_1}(q_2,q_1;E)&
A_{31;SI;\Lambda N(\Lambda)}^{ns_2i_2rs_1i_1}(q_2,q_1;E)\cr}\right),
\label{gl3}
\end{equation}

\begin{equation}
A_{13;SI}^{rs_1i_1ms_3i_3}(q_1,q_3;E)  =\left(\matrix{
A_{13;SI;N\Sigma}^{rs_1i_1ms_3i_3}(q_1,q_3;E)  &
   0 \cr
   0&
A_{13;SI;N\Lambda}^{rs_1i_1ms_3i_3}(q_1,q_3;E)  \cr}\right) \, ,
\label{gl4}
\end{equation}
where

\begin{eqnarray}
A_{23;SI;\alpha\beta}^{ns_{2}i_{2}ms_{3}i_{3}}(q_{2},q_{3};E)
&=&h_{23;SI}^{s_{2}i_{2}s_{3}i_{3}}\sum_{r}\tau
_{2;s_{2}i_{2};\alpha\beta}^{nr}(E-q_{2}^{2}/2\nu _{2}){\frac{q_{3}^{2}}{2}}  \nonumber \\
&&\times \int_{-1}^{1}d{\rm cos}\theta \,{\frac{P_{r}(x^\prime_{2})P_{m}(x_{3})}{%
E+\Delta E\delta_{\beta\Lambda}-p_{3}^{2}/2\mu _{3}-q_{3}^{2}/2\nu _{3}
+ i\epsilon}}; \,\,\,\,\,\,\,\, \alpha,\beta=\Sigma,\Lambda,
\label{gl5}
\end{eqnarray}

\begin{eqnarray}
A_{31;SI;\alpha N(\beta)}^{ns_{2}i_{2}ms_{1}i_{1}}(q_{2},q_{1};E)
&=&h_{31;SI}^{s_{2}i_{2}s_{1}i_{1}}\sum_{r}\tau
_{3;s_{2}i_{2;\alpha\beta}}^{nr}(E-q_{2}^{2}/2\nu _{2})
{\frac{q_{1}^{2}}{2}}  \nonumber \\
&&\times \int_{-1}^{1}d{\rm cos}\theta \,{\frac{P_{r}(x^\prime_{3})P_{m}(x_{1})}{%
E+\Delta E\delta_{\beta\Lambda}-p_{1}^{2}/2\mu _{1}-q_{1}^{2}/2\nu _{1}
+ i\epsilon}}; \,\,\,\,\,\,\,\, \alpha,\beta=\Sigma,\Lambda,
\label{gl6}
\end{eqnarray}

\begin{eqnarray}
A_{13;SI;N\beta}^{ns_{1}i_{1}ms_{3}i_{3}}(q_{1},q_{3};E)
&=&h_{13;SI}^{s_{1}i_{1}s_{3}i_{3}}\sum_{r}\tau
_{1;s_{1}i_{1;NN}}^{nr}(E+\Delta E\delta_{\beta\Lambda}-q_{1}^{2}/2\nu _{1})
{\frac{q_{3}^{2}}{2}}  \nonumber \\
&&\times \int_{-1}^{1}d{\rm cos}\theta \,{\frac{P_{r}(x^\prime_{1})P_{m}(x_{3})}{%
E+\Delta E\delta_{\beta\Lambda}-p_{3}^{2}/2\mu _{3}-q_{3}^{2}/2\nu _{3}.  
+ i\epsilon}}; \,\,\,\,\,\,\,\, \beta=\Sigma,\Lambda,
\label{gl7}
\end{eqnarray}
with the isospin and mass of particle 1 (the hyperon) being 
determined by the subindex $\beta$. $\mu_i$ and $\nu_i$ are the usual
reduced masses and the subindex $\alpha N(\beta)$ in 
Eq. (\ref{gl6}) indicates a transition $\alpha N \to \beta N$ with a
nucleon as spectator followed by a $NN \to NN$ transition with $\beta$ as
spectator. 
$\tau_{2;s_2i_2;\alpha\beta}^{nr}(e)$ 
are the coefficients of the 
expansion in terms of Legendre polynomials of the hyperon-nucleon
$t-$matrix $t_{2;s_2i_2;\alpha\beta}(p_2,p_2^\prime;e)$ for the 
transition $\alpha N \to \beta N$, i.e.,

\begin{equation}
\tau_{i;s_ii_i;\alpha \beta}^{nr}(e)={2n+1 \over 2}\,{2r+1 \over 2}
\int_{-1}^1 dx_i \int_{-1}^1
dx^\prime_i\, P_n(x_i)t_{i;s_ii_i;\alpha \beta}(p_i,p^\prime_i;e)
P_r(x^\prime_i) \, .
\label{for112}
\end{equation}

The energy shift $\Delta E$, which is usually taken as $M_\alpha-M_\beta$,
will be chosen instead such that at the $\beta d$ threshold the momentum
of the $\alpha d$ system has the correct value,
i.e., 

\begin{equation}
\Delta E={[(m_\beta+m_d)^2-(m_\alpha+m_d)^2]
[(m_\beta+m_d)^2-(m_\alpha-m_d)^2]\over
8\mu_{\alpha d}(m_\beta+m_d)^2},
\label{gl11}
\end{equation}
where $\mu_{\alpha d}$ is the $\alpha d$ reduced mass.

The inhomogeneous term of Eq. (\ref{for20}), 
$B_{2;SI;\beta}^{ns_2i_2}(q_2)$ is a two-component vector

\begin{equation}
B_{2;SI;\beta}^{ns_2i_2}(q_2) = \left( \matrix{
B_{2;SI;\Sigma\beta}^{ns_2i_2}(q_2)  \cr
B_{2;SI;\Lambda\beta}^{ns_2i_2}(q_2) \cr } \right),
\label{for3}
\end{equation}
where

\begin{equation}
B_{2;SI;\alpha\beta}^{ns_2i_2}(q_2)=h_{31;SI}^{s_2i_210}\phi_d(q_2)
\sum_r \tau_{2;s_2i_2;\alpha\beta}^{nr}(E_\beta^{th}-q_2^2/2\nu_2)
P_r(x_2^\prime).
\label{for4}
\end{equation}
$h_{31;SI}^{s_2i_2s_1i_1}$ with $s_1=1$ and $i_1=0$ are the spin-isospin
transition coefficients corresponding to a hyperon-deuteron initial state
(see Eq. (30), of Ref. \cite{Ter06}), $\phi_d(q_2)$ is the deuteron wave
function, $E_\beta^{th}$ is the energy of the $\beta d$ threshold,
$P_r(x_2^\prime)$ is a Legendre polynomial
of order $r$, and

\begin{equation}
x_2^\prime={{\eta_2\over m_3}q_2 - b\over {\eta_2\over m_3}q_2 + b}.
\label{for5}
\end{equation}

Finally, after solving the inhomogeneous set of equations (\ref{for20}),
the $\beta d$ scattering length is given by

\begin{equation}
A_{\beta d}=-\pi\mu_{\beta d}T_{\beta\beta},
\label{for6}
\end{equation}
with

\begin{equation}
T_{\beta\beta}=2\sum_{ns_2i_2}h_{13;SI}^{10s_2i_2}\int_0^\infty
q_2^2 dq_2 \phi_d(q_2)P_n(x_2^\prime) T_{2;SI;\beta\beta}^{ns_2i_2}(q_2).
\label{for7}
\end{equation}
In the case of the $\Sigma NN$ system,
even for energies below the $\Sigma d$ threshold, 
one encounters the three-body singularities of the
$\Lambda NN$ system so that to solve the integral equations (\ref{for20})
one has to use the contour rotation method where the momenta are 
rotated into the complex plane $q_i\to q_i e^{-i\phi}$ since as pointed
out in Ref. \cite{Ter06} the results do not depend on the contour rotation
angle $\phi$. 

\section{Results}
\label{res}

In order to solve the integral equations (\ref{for20}) for the
coupled $\Sigma NN - \Lambda NN$ system we consider all configurations
where the baryon-baryon subsystems are in an $S-$wave and the third 
particle is also in an $S-$wave with respect to the pair. 
However, to construct the two-body $t-$matrices that serve as
input of the Faddeev equations we considered the
full interaction including the contribution
of the $D-$waves and of course the coupling between the $\Sigma N$ and
$\Lambda N$ subsystems (this is known as the truncated $t-$matrix
approximation \cite{Ber86}). We give in Table \ref{t0c} the two-body channels
that are included in our calculation. For a given three-body state
$(I,J)$ the number of two-body channels that enter is determined by
the triangle selection rules $|J-{1\over 2}| \le s_i \le J+{1\over 2}$ 
and $|I-{1\over 2}| \le i_i \le I+{1\over 2}$. For the parameter
$b$ in Eqs. (8) and (9) we found that $b=3$ fm$^{-1}$ leads to very
stable results while for the expansion (7) we took twelve Legendre polynomials,
i.e., $0\le n \le 11$.

The results of our model correspond to the first
line of Tables \ref{t1}, \ref{t2}, and \ref{t3}.
In order to study their dependence on the
strength of the spin-singlet and spin-triplet 
hyperon-nucleon interactions,
we will construct different families of interacting potentials by introducing 
small variations of the parameters of the effective scalar exchange potential
around our reference values given in Table \ref{t0b}. These families 
are resumed in Table \ref{t1} and they are characterized by
their $\Lambda N$, $a_{i,s}$, and $\Sigma N$, $a_{i,s}^\prime$,
scattering lengths.
The $\Sigma N$ scattering lengths $a_{1/2,s}^\prime$ are 
complex since for these quantum numbers the 
$\Lambda N$ channel is always open.
Under the first row of Table IV, we give in parenthesis the
scattering length of the hyperon-nucleon channels as reproduced
by the expansion (7) in order to see how much they deviate from
the exact values. As one can see, the deviations are typically
of the order of 1\%.
We show in Fig. \ref{f1} the cross sections near threshold of the five
hyperon-nucleon processes for which data are available and compare
them with the predictions of our model with the 
scalar mesons taken as indicated in Table \ref{t0b} (solid line). 
The dashed lines stand for the results with the smallest
values for the spin-triplet scattering lengths 
$a_{1/2,1}$=1.41 fm and
$a'_{1/2,1}$=2.74 + $i$ 1.22 fm.
As can be seen when the spin-triplet scattering lengths decrease
the description of the experimental scattering cross sections 
is worsened.

Let us first comment on some relevant details about the two-body interactions
already reflected in the scattering lengths of Table \ref{t1}.
The $^1S_0$ $\Lambda N$ interaction becomes slightly more attractive than
the $^3S_1$ due to the scalar meson exchanges, because the OGE
is more repulsive for the spin-triplet than for the spin-singlet
partial wave. This is a desirable effect of the $^1S_0$ and $^3S_1$ $\Lambda N$
interactions, because the phase shifts around $p_\Lambda=$ 200 MeV/c for few-body
calculations of $s-$shell $\Lambda-$hypernuclei are expected to satisfy
$\delta(^1S_0) - \delta(^3S_1) \leq$ 10$^o$ \cite{Miy99}. It is also
interesting to note the strong repulsion appearing in the $\Sigma N$ system
for $(I,J)=(0,1/2)$ and $(1,3/2)$ $L=0$ partial waves. 
This is a consequence of the presence of an almost forbidden 
state due to the same symmetry of the spin and isospin wave functions.
This is clearly reflected in Table \ref{t1} in the small value
and negative sign of the $a_{3/2,1}^\prime$ and $a_{1/2,0}^\prime$
$\Sigma N$ scattering lengths as compared to the positive sign and large
value of the attractive $a_{3/2,0}^\prime$ and $a_{1/2,1}^\prime$
$\Sigma N$ scattering lengths. The importance of the $\Lambda N \leftrightarrow
\Sigma N$ potential for the $\Lambda NN$ and $\Sigma NN$ systems
will be discussed at the end of this section.
 
\subsection{The $\Lambda NN$ system}

Among the four possible $S-$wave positive parity states, $J=1/2$ and $3/2$
with isospin $I=0$ and $I=1$, those with isospin zero are the more
interesting ones. The $I=0$, $J^P=1/2^+$ is the channel where the
hypertriton lies, while the $I=0$, $J^P=3/2^+$ seems to be also
attractive. The possibility whether
states of total isospin and total angular momentum other than
$(I,J)=(0,1/2)$ could be also bound \cite{Miy95} will be 
investigated by studying the most natural candidates
$(I,J)=(0,3/2)$, $(1,1/2)$, and $(1,3/2)$.

The very weak binding energy of the hypertriton leads
one to expect that the wave function is mostly a deuteron surrounded
by a distant $\Lambda$ particle. As a consequence, since the $\Lambda$ 
particle is far apart from the two-nucleon subsystem, the on-shell
properties of the $\Lambda N$ and $\Sigma N$ interactions are expected
to be well reflected. In particular, this system is very well suited to
learn about the relative strength of the $^1S_0$ and $^3S_1$
attraction of the $\Lambda N$ interaction. The $^1S_0$ component
plays a more important role than the $^3S_1$ in the hypertriton,
although the available low-energy $\Lambda p$ total cross section
data cannot discriminate among the different combinations of the two
$S-$wave interactions. 

Let us first present the results for the isospin-0 channels.
We show in Table \ref{t2} the results obtained from the various 
cases of Table \ref{t1}. For the channel with total spin 3/2 we 
give the scattering length, $A_{0,3/2}$, while for the channel 
with total spin 1/2 we give the low-energy parameters,
$A_{0,1/2}$ and $R_{0,1/2}$, as well as the hypertriton
binding energy, $B_{0,1/2}$. Since the binding energy is very small the 
effective range, $R_{0,1/2}$, has been obtained from the binding energy 
and the scattering length using the relation,

\begin{equation}
R_{0,1/2} =(\sqrt{2\mu_{\Lambda d}B_{0,1/2}}
+1/A_{0,1/2})/(\mu_{\Lambda d}B_{0,1/2}).
\label{for8}
\end{equation}

\noindent
We observe how the increase of the three-body system binding energy
implies that the effective range of the interaction becomes
smaller, the essence of the variational argument made in Ref. \cite{Tho35}
to show that the nuclear force must have a finite (nonzero) range
or the triton would collapse to a point.

As one can see in Table \ref{t2} the values for 
the spin-3/2 scattering length, $A_{0,3/2}$, 
are very large, which means that there is a pole very
near threshold. Moreover, in some cases it becomes negative.
This would imply that the pole is a bound state
of which there is no experimental evidence whatsoever.
Since in the $(I,J)=(0,3/2)$ channel the hyperon-nucleon interaction 
with spin-singlet does not contribute this means those cases producing
a (0,3/2) bound state have too much attraction in the spin-triplet
hyperon-nucleon interaction. This can be easily checked in Table \ref{t1}
noting that the bound state appears when increasing 
the $a_{1/2,1}$ scattering length, the result being
independent of the $a_{1/2,0}$ scattering length. 
We plot in Fig. \ref{f2} the inverse of the
$\Lambda d$ scattering length, $1/A_{0,3/2}$, as a function of the 
$\Lambda N$ spin-triplet scattering length, $a_{1/2,1}$, for all 
cases in Table \ref{t1}, which as one sees falls within a 
smooth curve. From this figure one sees that a bound state will
appear if $a_{1/2,1}$ is larger than 1.68 fm. Thus,
if we impose the condition that the $(I,J)=(0,3/2)$ bound state should 
not appear we find that the scattering
length of the $\Lambda N$ $(i,j)=(1/2,1)$ channel must be smaller
that 1.68 fm. Moreover, we found in Fig. \ref{f1} that the fit 
of the hyperon-nucleon cross sections is worsened 
for those cases where the spin-triplet $\Lambda N$
scattering length is smaller than 1.41 fm,
so that we conclude that $1.41\le a_{1/2,1} \le 1.68$ fm.
From Table \ref{t1} one can estimate ranges of 
validity for the other $\Lambda N$ and $\Sigma N$
scattering lengths.

The spin 1/2 scattering length, $A_{0,1/2}$, is negative 
since there is a bound
state in this channel (the hypertriton). The experimental
value of the hypertriton binding energy is 0.13 $\pm$ 0.05 MeV and, as
it can be seen in Table \ref{t2}, the results of all models 
are consistent with the experimental value. 
We plot in Fig. \ref{f3} the low-energy parameters 
$R_{0,1/2}$ and $A_{0,1/2}$ as a function of the 
hypertriton binding energy for the different models 
of Table \ref{t1} where as one can see they both fall
within smooth curves. Our model gives a hypertriton
binding energy of 0.124 MeV, and the
low-energy parameters have the values
$R_{0,1/2}=3.82$ fm and $A_{0,1/2}=-17.2$ fm.

Finally, let us consider the isospin-1 channels.
We show in Fig. \ref{f4} the Fredholm
determinant of the $(I,J)=(1,1/2)$ and $(1,3/2)$
channels for energies below the $\Lambda NN$
threshold. The channel $(1,1/2)$ is attractive but not enough to
produce a bound state (i.e., the Fredholm determinant does not cross 
the negative real axis) while the channel (1,3/2) is repulsive.
The results for the other cases in Table \ref{t1}
are very similar to those of Fig. \ref{f4}. Let us finally
say that the $\Lambda NN$ channels could be ordered by the
strength of their attraction in the following way
$(0,1/2) \ge (0,3/2) \ge (1,1/2) \ge (1,3/2)$. A similar
conclusion was obtained in the simultaneous study
of the $\Lambda (\Sigma) NN$ systems of Ref. \cite{Miy95}
using the Nijmegen hyperon-nucleon and realistic $NN$ 
interactions.

Considering altogether the $J=1/2$ and $J=3/2$
$\Lambda NN$ channels, information about the 
relative contribution of the $^1S_0$
and $^3S_1 - ^3D_1$ hyperon-nucleon forces 
can be obtained.
As explained above the nonexistence of a $J=3/2$ $\Lambda NN$
bound state provides with an upper limit to the value of 
the spin-triplet $\Lambda N$ scattering length. The remaining
contribution to the hypertriton binding energy comes from
the spin-singlet partial wave. When the hypertriton is 
correctly described,
the contribution to the $\Lambda p$ scattering cross section
of the spin-singlet and spin-triplet partial waves is 
comparable and they are not
in the ratio $3:1$ as could be expected. This ratio 
for their relative contribution would either overestimate
the $\Lambda p$ scattering cross section or underestimate
the hypertriton binding energy.

\subsection{The $\Sigma NN$ system}

While the existence of strangeness $-1$ $\Lambda -$hypernuclei is well
established from the observation of many bound states, such has not been
the case for $\Sigma -$hypernuclei. In the late 80's experiments on
the recoilless production of $p-$shell hypernuclei \cite{Yam85}
seemed to indicate the existence of 
structures that might correspond to unbound states near threshold. 
Higher statistic experiments were not able to reproduce those
earlier peaks, concluding that they were fluctuations in poor 
statistics \cite{Bar99}.
However, there had been a KEK $K^- \to \pi^-$ at rest experiment
on $^4{\rm He}$ giving a small $^4_\Sigma{\rm He}$ peak \cite{Hay89}.
In flight $K^- \to \pi^-$ experiments at 600 MeV/c 
with high statistics \cite{Nag98} found that this 
peak was more convincingly
prominent, establishing this one $^4_\Sigma {\rm He}$ 
$\Sigma -$hypernucleus. The possible existence of bound states
or resonances of $\Sigma$'s in nuclei has become one of the main 
objectives of the FINUDA Collaboration in DAFNE \cite{Pia02} as well
as the KEK laboratory \cite{Suz04}, although the interpretation
of the results is under discussion \cite{Ose06}.

We show in Table \ref{t3} the $\Sigma d$ scattering lengths 
$A_{1,3/2}^\prime$ and $A_{1,1/2}^\prime$ for all cases in Table \ref{t1}.
The $\Sigma d$ scattering lengths are 
complex since the inelastic $\Lambda NN$ channels are always open.
The scattering length $A_{1,3/2}^\prime$ depends only on the
spin-triplet hyperon-nucleon channels and both its real and
imaginary parts increase when the spin-triplet hyperon-nucleon
scattering length increases. Our model, reproducing the
experimental hypertriton binding energy, gives 
$A_{1,3/2}^\prime = 0.36+i\,\,0.29$ fm.
The scattering length $A_{1,1/2}^\prime$ has a negative real part
which means that there is a quasibound state very near the 
$\Sigma d$ threshold in the $(I,J)=(1,1/2)$ channel.
The possible existence of a $\Sigma NN$ hypertriton
bound state or resonance has already being discussed in the literature \cite{Afn93}
and the actual possibilities for the experimental study at KEK
and FINUDA of this system claims for an experimental effort to
look for its possible existence.

We show in Fig. \ref{f5} the real part of the Fredholm determinant of the six
$(I,J)$ $\Sigma NN$ channels that are possible for energies below the 
$\Sigma d$ threshold. The imaginary part of the Fredholm determinant
is small and uninteresting. As one can see the channel (1,1/2)   
is the most attractive one since the Fredholm determinant is close to zero
at the $\Sigma d$ threshold, which as mentioned before,
indicates the presence of a quasibound state.
The next channel, in what to amount
of attraction is concerned, is the (0,1/2). 
The states generated by our model have widths of about 0.5 MeV.

Let us make a brief comment on the comparison with our previous
results for the $\Sigma NN$ system published in Ref. \cite{Ter06}.
In that work we only worried about the $\Sigma NN$ system and
we adjusted the experimental cross sections using the same
scalar exchange for the different spin channels, a naive 
generalization of the simple $SU(2)$ linear realization 
of chiral symmetry. The amount of attraction was controlled
by means of the harmonic oscillator parameter of the 
strange quark, $b_s=0.7$ fm. This procedure does modify
all the interactions simultaneously, gluons, pions, etc. 
increasing the amount of attraction in some cases by reducing
the global strength of the repulsive gluon or pion exchanges.
The inclusion of the $SU(3)$ scalar mesons gives a different
strength for the attraction of the different spin channels
and allows to use a more conventional harmonic oscillator
parameter for the strange quark, $b_s=0.55$ fm. Our former
calculation had therefore a too attractive spin-singlet 
hyperon-nucleon interaction and a too repulsive spin-triplet.
When studying the $\Lambda NN$ system it would give a reasonable
hypertriton binding energy but a too large $\Lambda p$ elastic cross
section. 
This detailed balance between the spin-singlet and
spin-triplet hyperon-nucleon strength, provided naively 
by the $SU(3)$ scalar quark-meson exchange (see Table \ref{t0b}),
can only be noticed when the full set of observables of the
different $\Lambda (\Sigma) NN$ systems are simultaneously
considered. Regarding the energy ordering of the different
$\Sigma NN$ $J=1/2$ isospin channels, our arguments are
fully maintained. The order of the two
attractive channels can be easily understood by looking at Table III.
All the attractive two-body channels 
in the $NN$ and $\Sigma N$
subsystems contribute to the $(I,J)=(1,1/2)$ $\Sigma NN$ state (the $\Sigma N
$ channels $^{3}S_{1}(I=1/2)$ and $^{1}S_{0}(I=3/2)$ and the $^{3}S_{1}(I=0)$
$NN$ channel), while the $(I,J)=(0,1/2)$ state does not present contribution
from two of them, the $^{1}S_{0}(I=3/2)$ $\Sigma N$ and specially the
$^{3}S_{1}(I=0)$ $NN$  deuteron channel. Actually, the $NN$ deuteron-like
contribution plays an essential role in the binding of the triton \cite{Val05}
and hypertriton \cite{Miy95}. In this last case the presence of the $\Lambda$
has the effect of reducing the $NN$ attraction with respect to the deuteron
case but the $\Lambda \leftrightarrow \Sigma$ conversion compensates 
this reduction and binds the system as we will see later on.

\subsection{$\Lambda \leftrightarrow \Sigma$ conversion}

It has been pointed out the relevance of small admixture of 
$\Sigma NN$ components to bind the hypertriton \cite{Miy95}
and also for other observables in hypernuclear physics \cite{Nog02}.
The contribution of the $\Lambda \leftrightarrow \Sigma$ conversion
should be even stronger than $\Delta \leftrightarrow N$ conversion
in ordinary nuclei, because it is not suppressed in $S-$waves and
the $\Lambda - \Sigma$ mass difference is much smaller. We have
investigated the effect of the $\Lambda \leftrightarrow \Sigma$
conversion both in the $\Lambda NN$ and in the $\Sigma NN$
systems. For this purpose we have solved the most
interesting $\Lambda NN$ channels, $(I,J)=(0,1/2)$ and $(0,3/2)$,
switching off the transition between 
the $\Lambda N$ and $\Sigma N$ subsystems. 
This excludes $\Sigma NN$ states between consecutive
$t$ operations and one stays therefore always in the space
of $\Lambda NN$ states. This intermediate transitions
from $\Sigma$ to $\Lambda$ induce a three-body force together with
a dispersive effect. 
We plot in Fig. \ref{f6}(a) the Fredholm determinant for both cases.
The solid line indicates the result of the full calculation while the
dashed one represents the results without $\Lambda \leftrightarrow \Sigma$
conversion. As can be seen
the effect of the $\Sigma NN$ channel for $\Lambda NN$
is very important. In fact the hypertriton bound state disappears
and the ordering between the $J=1/2$ and $J=3/2$ channels is reversed.

We have used the same strategy to evaluate the effect of $\Lambda NN$
channels for the $\Sigma NN$ system. We have redone the calculation
for the most interesting $\Sigma NN$ channels,
$(I,J)=(0,1/2)$ and $(1,1/2)$, disconnecting the 
$\Sigma \leftrightarrow \Lambda$ conversion. 
The results are shown in Fig. \ref{f6}(b). 
Once again the solid lines represent the full
calculation and the dashed ones those without 
$\Sigma \leftrightarrow \Lambda$ conversion. 
Once the transition between the $\Sigma N$ 
and $\Lambda N$ channels is disconnected
neither the character nor the order of these channels is modified. 
This is in agreement with our former conclusion that the inclusion 
of $\Lambda NN$ channels in the calculation of the $\Sigma NN$
system did not modify the general trend of the results \cite{Ter06}.

To firmly establish the relative importance of the conversion we
compare in Fig. \ref{f6}(c) the role of the $\Sigma \leftrightarrow \Lambda$
coupling in the same $(I,J)$ channel, in particular in the channel of the
hypertriton $(I,J)=(0,1/2)$. Some precaution is needed with respect
to the $x$-axis. The results are drawn as a function of the
energy below threshold, those of the $\Lambda NN$ system being
referred to the $\Lambda NN$ threshold and those of the $\Sigma NN$
system to the $\Sigma NN$ threshold.
As can be seen the effect of the lower $\Lambda NN$ channel
is less important for the heavier one, $\Sigma NN$.

\section{Summary}
\label{sum}

We have solved the Faddeev equations for the $\Lambda NN$ and
$\Sigma NN$ systems using the hyperon-nucleon and nucleon-nucleon
interactions derived from a chiral constituent quark model with 
full inclusion of the $\Lambda \leftrightarrow \Sigma$ conversion.
We present results for the hypertriton binding energy and the
$\Lambda d$ and $\Sigma d$ scattering lengths.
We also have calculated the elastic and
inelastic $\Sigma N$ and $\Lambda N$ cross sections.
We study consistently the $\Sigma NN$ system at threshold.

The set of observables studied gives rise to well-defined intervals 
for the possible values of the two-body $\Lambda N$ and $\Sigma N$
scattering lengths. The presence of almost forbidden states in the
$\Sigma N$ system produces observable effects for the sign and 
magnitude of the $\Sigma N$ scattering lengths.

The hypertriton turns out to be bound at the experimental binding
energy. We have found that the $\Lambda \leftrightarrow \Sigma$
conversion is crucial for the binding of the hypertriton. It is
also of interest to note that the contribution of the $^1S_0$ and
$^3S_1 - ^3D_1$ hyperon-nucleon forces to the binding are not
in the ratio $3:1$ as could be expected, but they are comparable.
We also found that the flavor dependence of the scalar meson-exchange
between quarks, giving rise to different strengths for the $^1S_0$
and $^3S_1 - ^3D_1$ attraction, becomes crucial for the understanding
of the full set of observables studied. From the nonexistence of a 
$\Lambda NN$ $(I,J)=(0,3/2)$ bound state, we derive an upper limit 
for the spin-triplet $\Lambda N$ scattering length, and a lower
limit can be derived from the correct description of two-body 
scattering cross sections. 

The $\Sigma NN$ system does not present any bound state. The 
$(I,J)=(1,1/2)$ and $(I,J)=(0,1/2)$ states are the only attractive
$S-$wave channels, the $(I,J)=(1,1/2)$ with a quasibound state or
resonance close to the three-body threshold. The channel with $I=1$
is always more attractive than the one with $I=0$. 
We have also studied the effect
of the $\Lambda \leftrightarrow \Sigma$ conversion for the $\Sigma NN$
system, its effect being much smaller that in the $\Lambda NN$ channel.

\acknowledgements
It is a great pleasure to thank Pedro Gonz\'alez for a careful reading of the 
manuscript and challenging and fruitful discussions as well as many
suggestions about this work.
This work has been partially funded by Ministerio de Educaci\'{o}n y Ciencia
under Contract No. FPA2004-05616, by Junta de Castilla y Le\'{o}n
under Contract No. SA-104/04, by Generalitat Valenciana under
Contract No. GV05/276, and by COFAA-IPN (M\'{e}xico).

\begin{table}[tbp]
\caption{Expectation value of the different 
scalar meson-exchange flavor operators.}
\label{t0}
\begin{tabular}{|c|cc|cc|cc|cc|}
 & \multicolumn{2}{c|}{$<N\Lambda | O_{ij} | N\Lambda >$} & 
\multicolumn{4}{c|}{$<N\Sigma | O_{ij} | N\Sigma >$} & 
\multicolumn{2}{c|}{$<N\Lambda | O_{ij} |N\Sigma >$} \\
\cline {2-9}
 $O_{ij}$ & \multicolumn{2}{c|}{$I=1/2$} &
 \multicolumn{2}{c|}{$I=1/2$} &
 \multicolumn{2}{c|}{$I=3/2$} &
 \multicolumn{2}{c|}{$I=1/2$} \\
\cline{2-9}
 & $S=0$ & $S=1$ & $S=0$ & $S=1$ & $S=0$ & $S=1$ & $S=0$ & $S=1$ \\
\hline
$\sum_{F=1}^{3} \lambda^F_3 \cdot \lambda^F_6$ & 0 & 0 & $-$4/9 & $-$4/9 
& 2/9  & 2/9 & 0 & 0 \\ 
$\sum_{F=4}^{7} \lambda^F_3 \cdot \lambda^F_6$ & 1/3 & $-$1/3 & $-$1/9 & 1/9 
& 2/9  & $-$2/9 & 1/3 & $-$1/3 \\ 
$\lambda^8_3 \cdot \lambda^8_6$ & 0 & 0 & 0 & 0 & 0 & 0 & 0 &0 \\ 
$\lambda^0_3 \cdot \lambda^0_6$ & 2/3 & 2/3 & 2/3  & 2/3 & 2/3 & 2/3 & 2/3 & 2/3 \\ 
\end{tabular}
\end{table}

\begin{table}[tbp]
\caption{Parameters, in fm$^{-1}$, of the $I=1/2$ scalar 
meson-exchange potential in Eq. (\ref{sca}).}
\label{t0b}
\begin{tabular}{ccccc}
 & Spin & $m_s$ & $\Lambda_{s}$ & \\
\cline{2-4 }
 & 0  & 3.73 & 4.2 &  \\ 
 & 1  & 4.12 & 5.2 &  \\ 
\end{tabular}
\end{table}

\begin{table}[tbp]
\caption{Two-body $\Sigma N$ channels $(i_\Sigma,s_\Sigma)$,
$\Lambda N$ channels $(i_\Lambda,s_\Lambda)$,
$NN$ channels with $\Sigma$ spectator $(i_{N(\Sigma)},s_{N(\Sigma)})$, and
$NN$ channels with $\Lambda$ spectator $(i_{N(\Lambda)},s_{N(\Lambda)})$ that contribute to
a given $\Sigma NN - \Lambda NN$ state with total isospin $I$ and spin $J$.}
\label{t0c}
\begin{tabular}{cccccc}
$I$ & $J$ & $(i_\Sigma,s_\Sigma)$  & $(i_\Lambda,s_\Lambda)$ 
& $(i_{N(\Sigma)},s_{N(\Sigma)})$ & $(i_{N(\Lambda)},s_{N(\Lambda)})$ \\ 
\tableline 
0 & 1/2 & (1/2,0),(1/2,1)  & (1/2,0),(1/2,1)  & (1,0) & (0,1) \\ 
1 & 1/2 & (1/2,0),(3/2,0),(1/2,1),(3/2,1)  & (1/2,0),(1/2,1) & (0,1),(1,0) &
(1,0)  \\ 
2 & 1/2 & (3/2,0),(3/2,1) & & (1,0) &  \\ 
0 & 3/2 & (1/2,1)  & (1/2,1) & & (0,1) \\ 
1 & 3/2 & (1/2,1),(3/2,1)  & (1/2,1) & (0,1) & \\ 
2 & 3/2 & (3/2,1) &  & & \\ 
\end{tabular}
\end{table}

\begin{table}[tbp]
\caption{$\Lambda N$ scattering lengths, $a_{1/2,0}$ and $a_{1/2,1}$,
and $\Sigma N$ scattering lengths, $a_{1/2,0}^\prime$, $a_{1/2,1}^\prime$,
$a_{3/2,0}^\prime$, and $a_{3/2,1}^\prime$ (in fm), dependence
on the strength of the spin-singlet and spin-triplet scalar interaction.
Masses are in fm$^{-1}$. The values in parenthesis under the first row
correspond to the scattering lengths obtained using the expansion (7)
with $0 \le n \le 11$.}
\label{t1}
\begin{tabular}{|cccccccc|}
 $m_s(S=0)$ & $m_s(S=1)$ &
$a_{1/2,0}$ & $a_{1/2,1}$ & 
$a_{1/2,0}^\prime$ & $a_{1/2,1}^\prime$ & 
$a_{3/2,0}^\prime$ & $a_{3/2,1}^\prime$ \\ 
\hline\hline 
3.73 & 4.12 & 2.48 & 1.65 & $-$0.66+$\, i\, $0.12 & 3.20+$\, i\, $1.52 & 3.90 & $-$0.42 \\ 
 & & (2.50) & (1.67) & ($-$0.64+$\, i\, $0.12) & (3.24+$\, i\, $1.52) & (3.96) & ($-$0.40) \\ \hline
3.76 & 4.12 & 2.31 & 1.65 & $-$0.66+$\, i\, $0.12 & 3.20+$\, i\, $1.52 & 3.64 & $-$0.42 \\ 
3.71 & 4.12 & 2.55 & 1.65 & $-$0.65+$\, i\, $0.12 & 3.20+$\, i\, $1.52 & 4.20 & $-$0.42 \\ 
3.68 & 4.12 & 2.74 & 1.65 & $-$0.65+$\, i\, $0.12 & 3.20+$\, i\, $1.52 & 4.72 & $-$0.42 \\ 
3.73 & 4.20 & 2.48 & 1.41 & $-$0.66+$\, i\, $0.12 & 2.74+$\, i\, $1.22 & 3.90 & $-$0.44 \\ 
3.73 & 4.10 & 2.48 & 1.72 & $-$0.66+$\, i\, $0.12 & 3.34+$\, i\, $1.62 & 3.90 & $-$0.42 \\ 
3.73 & 4.08 & 2.48 & 1.79 & $-$0.66+$\, i\, $0.12 & 3.48+$\, i\, $1.72 & 3.90 & $-$0.41 \\ 
3.73 & 4.06 & 2.48 & 1.87 & $-$0.66+$\, i\, $0.12 & 3.64+$\, i\, $1.85 & 3.90 & $-$0.41 \\ 
3.73 & 4.04 & 2.48 & 1.95 & $-$0.66+$\, i\, $0.12 & 3.81+$\, i\, $1.98 & 3.90 & $-$0.40 \\ 
\end{tabular}
\end{table}

\begin{table}[tbp]
\caption{$\Lambda d$ scattering lengths,
$A_{0,3/2}$ and $A_{0,1/2}$, effective range,
$R_{0,1/2}$ (in fm), and hypertriton binding energy
$B_{0,1/2}$ (in MeV) for all cases of Table \protect\ref{t1}}
\label{t2}
\begin{tabular}{|cccccc|}
$m_s(S=0)$ & $m_s(S=1)$ &
 $A_{0,3/2}$ & $A_{0,1/2}$ & $R_{0,1/2}$ & $B_{0,1/2}$ \\ 
\hline\hline
 3.73 & 4.12 & 198.2    & $-$17.2 & 3.82 & 0.124 \\ \hline
 3.76 & 4.12 & 198.2    & $-$22.4 & 4.47 & 0.070 \\ 
 3.71 & 4.12 & 198.2    & $-$16.8 & 3.80 & 0.130 \\ 
 3.68 & 4.12 & 198.2    & $-$14.4 & 3.51 & 0.182 \\ 
 3.73 & 4.20 &  66.3    & $-$20.0 & 4.17 & 0.089 \\ 
 3.73 & 4.10 & $-$179.8 & $-$16.6 & 3.75 & 0.134 \\ 
 3.73 & 4.08 & $-$62.7  & $-$16.0 & 3.68 & 0.145 \\ 
 3.73 & 4.06 & $-$38.2  & $-$15.4 & 3.61 & 0.156 \\ 
 3.73 & 4.04 & $-$27.6  & $-$14.9 & 3.55 & 0.168 \\ 
\end{tabular}
\end{table}

\narrowtext

\begin{table}[tbp]
\caption{$\Sigma d$ scattering lengths,
$A_{1,3/2}^\prime$ and $A_{1,1/2}^\prime$ (in fm),
for all cases in Table \protect\ref{t1}.}
\label{t3}
\begin{tabular}{|cccc|}
$m_s(S=0)$ & $m_s(S=1)$ & $A_{1,3/2}^\prime$ & $A_{1,1/2}^\prime$  \\ 
\cline{1-4}\cline{1-4}
3.73 & 4.12 & 0.36+$\, i\, $0.29 & $-$1.55+$\, i\, $42.31 \\ \cline{1-4}
3.76 & 4.12 & 0.36+$\, i\, $0.29 & 14.95+$\, i\, $31.61 \\ 
3.71 & 4.12 & 0.36+$\, i\, $0.29 & $-$21.04+$\, i\, $33.19 \\ 
3.68 & 4.12 & 0.36+$\, i\, $0.29 & $-$23.29+$\, i\, $13.32 \\ 
3.73 & 4.20 & 0.20+$\, i\, $0.26 & 19.28+$\, i\, $25.37 \\
3.73 & 4.10 & 0.40+$\, i\, $0.30 & $-$10.47+$\, i\, $40.25 \\ 
3.73 & 4.08 & 0.44+$\, i\, $0.31 & $-$17.33+$\, i\, $35.01 \\ 
3.73 & 4.06 & 0.49+$\, i\, $0.33 & $-$21.16+$\, i\, $28.54 \\ 
3.73 & 4.04 & 0.54+$\, i\, $0.34 & $-$22.44+$\, i\, $22.44 \\ 
\end{tabular}
\end{table}

\begin{figure}[tbp]
\caption{Calculated $\Lambda N$, $\Sigma N$ and $\Sigma N \rightarrow \Lambda N$
total cross sections compared with experimental data. 
Experimental data in (a) are from Ref. \protect\cite{Ale68}, 
in (b) and (c) are from Ref. \protect\cite{Eis71}, 
and in (d) and (e) from Ref. \protect\cite{Eng66}.}
\label{f1}
\end{figure}

\begin{figure}[tbp]
\caption{The inverse of the $(I,J)=(0,3/2)$ $\Lambda d$ scattering length
as a function of the $\Lambda N$ $a_{1/2,1}$ scattering length
for all models of Table \protect\ref{t1}. The solid line
is to guide the eye.}
\label{f2}
\end{figure}

\begin{figure}[tbp]
\caption{The $(I,J)=(0,1/2)$ $\Lambda d$ scattering length and effective range
parameters as a function of the hypertriton binding energy
for all models of Table \protect\ref{t1}. The solid line
is to guide the eye.}
\label{f3}
\end{figure}

\begin{figure}[tbp]
\caption{Fredholm determinant for  the $\Lambda NN$
channels $(I,J)$ = (1,1/2) and (1,3/2) for 
the model giving the solid line $\Sigma N$ total cross sections of Fig.
\protect\ref{f1} and 
for energies below the $\Lambda NN$ threshold.}
\label{f4}
\end{figure}

\begin{figure}[tbp]
\caption{Fredholm determinant for (a) $J=1/2$ and (b) $J=3/2$ $\Sigma NN$
channels for the model giving the solid line $\Sigma N$ total cross sections of Fig.
\protect\ref{f1}. The $\Sigma d$ continuum starts at $E=-2.225$ MeV, the
deuteron binding energy obtained within our model.}
\label{f5}
\end{figure}

\begin{figure}[tbp]
\caption{ (a) Fredholm determinant for $J=1/2$ and $J=3/2$ $I=0$ 
$\Lambda NN$ channels. The solid line stands for the full calculation
and the dashed line when the $\Lambda \leftrightarrow \Sigma$ transition
is taken to be zero. (b) Same as (a) for the $I=0$ and $I=1$ $J=1/2$
$\Sigma NN$ channels. (c) Same as (a) for $(I,J)=(0,1/2)$ $\Lambda NN$
and $\Sigma NN$ channels.}
\label{f6}
\end{figure}
\end{document}